\documentclass{article}
\usepackage[utf8]{inputenc}
\usepackage[english]{babel}
\usepackage{amsmath}
\usepackage{graphicx}
\usepackage{subcaption}
\usepackage{amssymb}
\usepackage{fixltx2e}
\usepackage{float}
\usepackage[margin=1.in]{geometry}
\usepackage{natbib}
\usepackage{chngcntr}
\usepackage{hyperref}
\hypersetup{
    colorlinks=true,
    linkcolor=blue,
    citecolor=blue,
}
\usepackage{cancel}
\usepackage[normalem]{ulem}
\usepackage{lscape}
\usepackage{rotating}
\usepackage[page]{appendix}
\usepackage[lined,boxed]{algorithm2e}

\usepackage{cleveref}
\usepackage{siunitx}
\usepackage{multirow}

\newcommand{\xt}{\tilde{x}}

\newcommand{\yt}{\tilde{y}}
\newcommand{\kt}{\tilde{k}}
\newcommand{\ttm}{\tilde{t}}

\newcommand{\tDt}{\tilde{D}_0}
\newcommand{\Lxt}{\tilde{L}_x}
\newcommand{\Lyt}{\tilde{L}_y}
\newcommand{\vxt}{\tilde{v}_x}

\newcommand{\vyt}{\tilde{v}_y}

\newcommand{\Ct}{\tilde{C}}

\newcommand{\Dsm}{D^*_{max}}
\newcommand{\Dsmopt}{\left(\Dsm\right)_{opt}}

\newcommand{\nablat}{\tilde{\nabla}}
\newcommand{\Bvt}{\tilde{{\mathbf v}}}

\newcommand{\Bxt}{\mathbf{\xt}}

\DeclareMathOperator*{\argmin}{argmin}

\title{A new flow-kinematics-based model for time-dependent effective dispersion in mixing-limited reactions}

\author{Ricardo H.~Deucher$^{a}$, Louis J.~Durlofsky$^a$}
\date {$^a$Department of Energy Resources Engineering, Stanford University, Stanford CA 94305 \\
\BlankLine
February 2022}
\begin{document}
\maketitle

\section*{Abstract}
A new upscaling procedure that provides 1D representations of 2D mixing-limited reactive transport systems is developed and applied. A key complication with upscaled models in this setting is that the procedure must differentiate between interface spreading, driven by the spatially variable velocity field, and mixing, in which components contact one another and react. Our model captures the enhanced mixing caused by spreading through use of a time-dependent effective dispersion term. The early-time behavior of this dispersion is driven by flow kinematics, while at late times it reaches a Taylor-dispersion-like limit. The early-time behavior is modeled here using a very fast (purely advective) particle tracking procedure, while late-time effects are estimated from scaling arguments. The only free parameter in the model is the asymptotic effective dispersion. This quantity is determined for a few cases by calibrating 1D results to reference 2D results. For most cases, it is estimated using a fit involving a dimensionless grouping of system variables. Numerical results for bimolecular reaction systems are generated using a pseudo-spectral approach capable of resolving fronts at high Peclet numbers. Results are presented for two different types of 2D velocity fields over a wide range of parameters. The upscaled model is shown to provide highly accurate results for conversion factor, along with reasonable approximations of the spatial distribution of reaction occurrence. The model is also shown to be valid for non-reacting systems, and results for such cases can be used in the calibration step to achieve computational savings.

\section{Introduction}

The accurate modeling of mixing-limited reactions is computationally challenging, as it requires the numerical resolution of sharp concentration gradients for multiple components in spatially variable velocity fields. The development of efficient numerical methods and accurate coarse-grained (upscaled) representations are essential steps for the modeling of mixing-limited reactions across scales. As the occurrence of the chemical reactions depends on the reactants contacting one another, the upscaling of such processes must differentiate between spreading and mixing. Spreading, in this context, is driven by flow kinematics, specifically by a spatially variable velocity field that distorts and spreads the interface region at which reactants interact. Mixing, which occurs in the interface region itself, brings the reactants into contact and allows them to react \citep{Valocchi2019}.

In this work, we introduce a time-dependent effective dispersion representation that captures the effects of spreading on mixing in upscaled (1D) models describing 2D reactive flow systems with spatially variable velocity fields. We consider fast, irreversible bimolecular reactions under advection-dominated conditions (i.e., at high Peclet number), though our treatments should be applicable to systems involving multiple reactive components. The governing equations, in both the reference 2D and upscaled 1D systems, are solved using a pseudo-spectral method that provides accurate representations of the sharp concentration gradients (which occur at large Peclet numbers) at reasonable computational cost \citep{adrover2002spectral}.

A wide range of studies have addressed mixing and mixing-limited reactions at the pore and Darcy scales. Our focus here is on Darcy-scale problems; for a discussion of pore-scale studies, please see the recent review by \cite{Valocchi2019}. \cite{Tartakovsky2009} and \cite{Battiato2011} investigated the conditions under which an upscaled representation of the classical advection-dispersion-reaction equation is appropriate for the description of mixing-limited reactions. They showed that the classical representation, with a constant dispersion coefficient, is only applicable for such modeling in a particular range of Peclet and Damkohler numbers. In the context of Darcy flow, the application of constant macroscopic dispersion coefficients to modeling mixing-limited reactions can lead to an overestimation of the reaction rate \citep{molz1988internal,kapoor1997bimolecular}, as these coefficients describe the combined effects of mixing and spreading. Within a stochastic modeling framework, \cite{dentz2000temporal} developed analytical expressions for a time-dependent effective dispersion coefficient that describes spreading from a point source. This model was derived under the assumptions of stationarity and small variance in log-conductivity. \cite{dentz2000temporal} pointed out that this effective dispersion coefficient more reliably represents physical mixing than the time-dependent macroscopic (or ensemble) dispersion coefficient, which describes the spreading of a large plume.  Recognizing the local nature of the chemical reactions, \cite{Cirpka2002} proposed to quantify mixing -- and not spreading -- through use of the time-dependent effective dispersion coefficients derived by \cite{dentz2000temporal} and \cite{fiori2000concentration}. \cite{Cirpka2002} showed that, with this approach, reaction rates were properly represented even though plume spreading was underestimated. In subsequent work, \cite{jose2004measurement} used the same concept to model column-scale reactive transport experiments.

The deformation of mixing fronts due to flow kinematics has been studied extensively within the context of turbulent and chaotic flow modeling \citep{ranz1979applications,ottino1989kinematics,duplat2008mixing}.
The lamella approach was developed by \cite{ranz1979applications} to quantify the interaction between fluid deformation (spreading) and mixing in laminar and turbulent flows. This treatment, implemented within a Lagrangian framework, establishes a link between the stretching along interfaces and mixing and reactions. As such, it provides a theoretical foundation for the quantification of mixing in spatially variable velocity fields. The geometry of the mixing interface has also been shown to control mixing and reaction rates at the pore \citep{DeAnna2014} and Darcy \citep{leBorgne2014impact} scales. The deformation of the mixing fronts increases their length and, as a consequence of mass conservation, enhances the concentration gradients in the direction orthogonal to the elongation. These mechanisms lead to enhanced mixing and reaction dynamics and form the basis of the upscaled model proposed in this work. 

To account for late-time aggregation between lamella (i.e., diffusive coalescence), \cite{villermaux2012mixing} and \cite{LeBorgne2015} considered a random aggregation process. \cite{perez2019upscaling} developed the dispersive lamella approach, which captures the effects of early-time fluid deformation and late-time dispersive mixing in a single mathematical framework. The dispersive lamella approach is based on an approximation of the Green's function for the advection-diffusion problem and considers a time-dependent effective dispersion coefficient that accounts for stretching enhanced diffusion at early-time and front coalescence at late time. This modeling framework was applied to the upscaling of mixing-limited reactions in Poiseuille flow \citep{perez2019upscaling}, in which case analytical expressions for the time-dependent effective dispersion coefficient in layered systems, derived by \cite{dentz2007mixing}, were employed. It has also been applied for pore-scale flows \citep{Perez2020, puyguiraud2020effective}. In this case the effective dispersion coefficient is based on the  width of the mixing interface as measured from experimental and/or detailed pore-scale simulations.

The upscaled model developed in this work is inspired by the studies discussed above, as it accounts for the effects of spreading on mixing and reaction through use of a time-dependent dispersion coefficient, $D^*(t)$. In contrast to some of the previous studies, we work within an Eulerian framework, and instead of estimating $D^*(t)$ from analytical expressions (as in \cite{Cirpka2002} and \cite{perez2019upscaling}), we construct the time-dependent portion of $D^*(t)$ from flow kinematics considerations. More specifically, the length of the mixing front is estimated under purely advective conditions using a very fast particle tracking procedure. To account for late-time effects, when diffusion counteracts the stretching imposed by the spatially variable velocity field, we propose a functional form for $D^*(t)$ that asymptotically approaches a Taylor-dispersion-like limit. This quantity, referred to as $\Dsm$, is the only parameter required by the upscaled model. Given reference 2D results, $\Dsm$ can be determined by solving a simple optimization problem. Through use of scaling arguments and a few `exact' $\Dsm$ values, we construct general fits that provide estimates of $\Dsm$ for new cases involving different velocity fields and/or Peclet numbers.

The paper is organized as follows. In Section~\ref{sec:govEquations}, the governing equations, dimensionless parameters, and velocity fields considered in this study are presented. In Section~\ref{sec:upscaledAll}, we describe the upscaled model and the time-dependent effective dispersion coefficient that captures the evolution of the mixing front. The overall workflow for applying our treatments to reactive transport problems is also provided. In Section~\ref{sec:applications}, we present results demonstrating the performance of the upscaled model for two different types of spatially variable 2D velocity fields. Conclusions and suggestions for future work in this area appear in Section~\ref{sec:conclusions}. Convergence results for the pseudo-spectral procedure used in this work are provided in an appendix.

\section{Governing equations} \label{sec:govEquations}

We consider the bimolecular reaction

\begin{equation}\label{eq:bimolecReaction}
    A + B \rightarrow C,
\end{equation}
where the components $A$ and $B$ are initially segregated, such that the reaction is driven by the mixing of components $A$ and $B$. For a given velocity field, the mass balance equations are

\begin{subequations}\label{eq:conserv_dim}
    \begin{align}
        \frac{\partial \Ct_i}{\partial \ttm} + \Bvt \cdot \nablat \Ct_i =  \tDt \nablat^2 \Ct_i - \kt \Ct_A \Ct_B   , \quad 0 < \xt < \Lxt, \quad 0 < \yt < \Lyt, \quad i = \textit{A }\text{and }\textit{B}, \label{eq:conservAB_dim} \\
        \frac{\partial \Ct_C}{\partial \ttm} + \Bvt \cdot \nablat \Ct_C =  \tDt \nablat^2 \Ct_C + \kt \Ct_A \Ct_B   , \quad 0 < \xt < \Lxt, \quad 0 < \yt < \Lyt, \label{eq:conservC_dim}
    \end{align}
\end{subequations}
where the tildes denote dimensional quantities, $\Ct_i$ is the concentration of component $i$ in \si{mole\cdot{} m^{-3}}, $\tDt$ is the constant isotropic diffusion/dispersion coefficient in \si{m^2 \cdot{} s^{-1}}, $\Bvt=\vxt{\mathbf i}_x+\vyt{\mathbf i}_y$ is the 2D incompressible velocity field ($\nablat \cdot \Bvt=0$), of units of \si{m \cdot{} s^{-1}}, $\kt$ is the kinetic rate constant in \si{m^3 \cdot mole^{-1} \cdot s^{-1}}, and $\Lxt$ and $\Lyt$ are the lengths of the domain in the $\xt$ and $\yt$ coordinate directions. Equation~\ref{eq:conserv_dim} describes either the reactive transport problem in a free fluid, with $\tilde{D}_0$ the diffusion coefficient, or reactive transport problem in porous media at the Darcy scale, in which case $\tilde{D}_0$ is the dispersion coefficient, taken here as a constant for simplicity \citep{leBorgne2014impact,Bandopadhyay2018,wright2017effects,nijjer2019stable}. Note that in the latter case, perfect mixing is assumed at the support scale.

The initial conditions are given in terms of a characteristic concentration $\Ct_0$ as

\begin{subequations}\label{eq:IC_dim}
    \begin{align}
        \Ct_A{(\Bxt,\ttm=0)} = 
        \begin{cases}
            \Ct_0 & \text{if }  \xt < \Lxt/2 \\
            0 & \text{otherwise} 
        \end{cases}, \\
        \Ct_B{(\Bxt,\ttm=0)} = 
        \begin{cases}
            \Ct_0 & \text{if }  \xt \geq \Lxt/2 \\
            0 & \text{otherwise} 
        \end{cases}, \\
        \Ct_C(\Bxt,\ttm=0) = 0.
\end{align}
\end{subequations}
where $\Bxt = (\tilde{x}, \tilde{y})$. Defining the average velocity in the main flow direction

\begin{equation}
    \bar{\tilde{v}}_x = \frac{1}{\Lyt}\int_{0}^{\Lyt} \vxt \,d\yt,
\end{equation}
and the convective time as $\ttm_c = \Lxt / \bar{\tilde{v}}_x$, we can introduce the following dimensionless quantities

\begin{equation} \label{eq:dimensionlessQuantities}
\begin{split}
    \mathbf{x} = \Bxt / \Lyt, \quad \mathbf{v} = \Bvt / \bar{\tilde{v}}_x,
    \quad t = \ttm / \ttm_c, \quad \lambda = \Lxt / \Lyt, \\
    C_i = \Ct_i / \Ct_0, \quad \text{for} \quad i = \textit{A, B} \text{, and } \textit{C}, \\
    Pe = \bar{\tilde{v}}_x \Lyt / \tilde{D}_0, \quad Da = \kt \Ct_0 \Lyt / \bar{\tilde{v}}_x.
\end{split}    
\end{equation}
The Peclet number $Pe$ represents the ratio of diffusion to advection time scales, and the advective Damk\"ohler number represents the ratio of advection to reaction time scales. Introducing the above nondimensionalizations, the dimensionless form of Eq.~\ref{eq:conserv_dim} is

\begin{subequations}\label{eq:conserv_adim}
    \begin{align}
        \frac{\partial C_i}{\partial t} + \mathbf{v} \cdot \nabla C_i =  \frac{1}{Pe} \nabla^2 C_i - Da C_A C_B, \quad 0 < x < \lambda, \quad 0 < y < 1, \quad i = \textit{A } \text{and } \textit{B}, \label{eq:conservAB_adim} \\
        \frac{\partial C_C}{\partial t} + \mathbf{v} \cdot \nabla C_C =  \frac{1}{Pe} \nabla^2 C_C + Da C_A C_B, \quad 0 < x < \lambda, \quad 0 < y < 1, \label{eq:conservC_adim}
    \end{align}
\end{subequations}
with initial conditions

\begin{subequations}\label{eq:IC_adim}
    \begin{align}
        C_A{(\mathbf{x},t=0)} = 
        \begin{cases}
            1 & \text{if }  x < \lambda/2 \\
            0 & \text{otherwise} 
        \end{cases}, \\
        C_B{(\mathbf{x},t=0)} = 
        \begin{cases}
            1 & \text{if }  x \geq \lambda/2 \\
            0 & \text{otherwise} 
        \end{cases}, \\
        C_C(\mathbf{x},t=0) = 0.
\end{align}
\end{subequations}

Although we are interested in general velocity fields $\mathbf{v} = \mathbf{v}(x,y)$, in this work we will consider two simplified analytical velocity fields, referred to as the oscillatory field and a modification of the ABC field \citep{ravu2016creating}. For the oscillatory field, $v_y=0$ and

\begin{equation}\label{eq:velFieldOsc}
    v_x = 1  + a \sin \left(2 \pi \omega y \right),
\end{equation}
where $a$ is the dimensionless amplitude of the velocity fluctuations and $\omega$ is the frequency of oscillation. For the ABC field

\begin{subequations}\label{eq:velFieldABC}
    \begin{align}
        v_x = 1 + \sum_{i}^{N_{f}} a_i \cos\left(2 \pi (\omega_i x + \gamma_i)\right) \cos\left(2 \pi (\omega_i y + \gamma_i)\right) \\
        v_y = \bar{v}_y + \sum_{i}^{N_{f}} a_i \sin\left(2 \pi (\omega_i x + \gamma_i)\right) \sin\left(2 \pi (\omega_i y + \gamma_i)\right), 
    \end{align}
\end{subequations}
where $N_{f}$ is the number of frequencies considered in the velocity distribution and $\gamma$ is a factor that acts to shift the alignment between different frequencies. Note that $a_i$ and $\omega_i$ need to be the same in the $v_x$ and $v_y$ expressions in order for the ABC field to maintain $\nabla \cdot \mathbf{v}=0$. In the context of porous media flows, the oscillatory velocity field corresponds to a layered permeability distribution with the pressure gradient in the $x$-direction. The ABC velocity field does not precisely correspond to a practical Darcy flow scenario, but it provides an analytical expression for the velocity with nonzero $x$ and $y$ components that can be used to test our treatments in a more complicated setting.

At the high $Pe$ and $Da$ numbers of interest here, the solution of Eq.~\ref{eq:conserv_adim} is highly sensitive to small amounts of numerical diffusion. Thus, numerical approaches based on finite-volume techniques require a very large number of grid blocks to achieve converged results. Following \cite{adrover2002spectral}, we solve Eq.~\ref{eq:conserv_adim} with a pseudo-spectral method that applies a Fourier approximation of the spatial derivatives and a fourth-order Runge-Kutta time integration. Periodic boundary conditions are imposed, but we only consider times prior to the point where periodic images of the concentration field interact \citep{Bandopadhyay2017}. We note that periodic boundary conditions have been widely used in subsurface flow modeling and upscaling. Specific examples include the computation of upscaled absolute permeability tensors \citep{Durlofsky1991} and modeling of mixing-limited reactions \citep{Bandopadhyay2017}.
\section{Upscaled model for mixing-limited reactions} \label{sec:upscaledAll}

As pointed out by \cite{dentz2007mixing}, reactions occur where (and when) reacting species contact one another. This means that mixing, and the mechanisms that influence mixing, are controlling factors for many reactive transport processes. \cite{deSimoni2005procedure} demonstrated that the reaction rate for mixing-limited reactions depends on the rate at which the components mix. For the special case of equilibrium reactions, they showed that the reactive transport problem can be reformulated as a transport equation for a conservative (i.e., non-reacting) component and an analytical expression for the reaction rate. This expression depends on the mixing rate of the conservative component and a reaction factor dependent on the speciation of the species involved in the equilibrium reactions. Thus we see that an understanding of mixing processes for conservative components has direct implications for mixing-limited reactive transport processes. Based on this insight, our model for the upscaling of mixing-limited reactions starts with the development of an upscaled model for the mixing of a conservative component. 

\subsection{Upscaled model for mixing} \label{sec:upscaledMixing}

We now develop an upscaled model for the advection-diffusion equation for a non-reactive component $\theta$. The 2D (fully resolved) mass balance equations reads

\begin{equation} \label{eq:mixing2D}
      \frac{\partial C_\theta}{\partial t} + \mathbf{v} \cdot \nabla C_\theta =  \frac{1}{Pe} \nabla^2 C_\theta, \quad 0 < x < \lambda, \quad 0 < y < 1.
\end{equation}
Our goal is to capture the key aspects of this 2D system with an upscaled 1D model of the form
\begin{equation} \label{eq:mixingUpscaled}
    \frac{\partial \bar{C}_\theta}{\partial t} +  \frac{\partial \bar{C}_\theta}{\partial x}= \frac{1}{Pe} D^*(t) \frac{\partial^2 \bar{C}_\theta}{\partial x^{2}} , \quad 0 < x < \lambda.
\end{equation}
Here $D^*(t)$ is the  dimensionless time-dependent effective dispersion coefficient (nondimensionalized by $\tilde{D}_0$), introduced to represent the complicated 2D mixing dynamics in the 1D representation. The overbars indicate averaged (1D) quantities.

\begin{figure} [H]
    \centering
    \includegraphics[width=0.99\textwidth,trim = 00 0 0 0,clip]{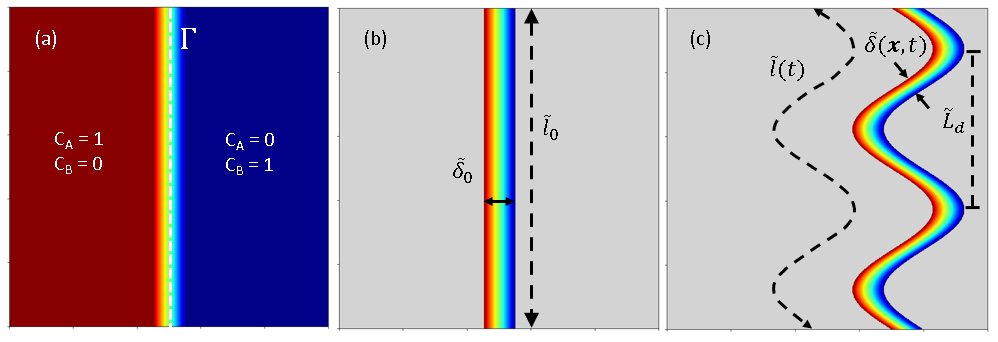}
    \caption{Mixing front stretching and deformation. (a) initial condition, smoothed for illustrative purposes; (b) highlight of the mixing front with initial length $\tilde{l}_0$, and initial width $\tilde{\delta}_0$; (c) mixing front deformation due to the spatially variable velocity field, with $\tilde{L}_d$ the distance between neighboring extrema.}
    \label{fig:stretch}
\end{figure}

The ideas behind our upscaled model are illustrated in Figure~\ref{fig:stretch}. In Figure~\ref{fig:stretch}a, the initial distribution of component $A$, given by Eq.~\ref{eq:IC_adim}a, is shown. For illustrative purposes, the mixing front ($\Gamma$) at $x = 0.5$ is depicted with a smooth concentration gradient and a finite thickness. In Figure~\ref{fig:stretch}b, the initial form of the mixing front is depicted. The front is of initial length $\tilde{l}_0$, and initial width $\tilde{\delta}_0$. The finite width ($\tilde{\delta}_0$) represented in Figure~\ref{fig:stretch}b is in a sense a hypothetical quantity, as the initial width of the mixing front in Eq.~\ref{eq:IC_adim} is zero. The finite width in Figure~\ref{fig:stretch}b can, however, be viewed as an early-time representation of a case with very large but finite $Pe$. As time proceeds, the length and width of the mixing front change due to the kinematics of the flow, as depicted in Figure~\ref{fig:stretch}c. Here $\tilde{L}_d$ represents the (dimensional) distance between neighboring extrema in the front.

We define $l = \tilde{l}/\Lyt$, $\delta = \tilde{\delta}/\Lyt$, and the average width as $\bar{\delta}(t) = \frac{1}{l(t)}\int_{\Gamma} \delta(\mathbf{x},t)d\Gamma$. Note that $l_0 = 1$ for the initial condition considered in this work. Under incompressible flow conditions and at early times, for which diffusive smearing of the mixing front is not yet significant, mass conservation requires that
\begin{equation}
    l(t) \bar{\delta}(t) = l_0 \delta_0 \equiv 1.
\end{equation}
Without loss of generality we set the initial width $\delta_0$ to be 1 (any value could be used for this), such that $l_0 \delta_0 \equiv 1$.

To properly represent the impacts of spreading on mixing, $D^*(t)$ should be determined such that the diffusive mass transfer rate across the mixing front, denoted by $J$, in the 1D representation (Eq.~\ref{eq:mixingUpscaled}) provides a reasonable approximation of the equivalent quantity in the 2D representation (Eq.~\ref{eq:mixing2D}). The 1D quantity, $J_{1D}$, can be expressed as
\begin{equation} \label{eq:1Dflux}
    J_{1D} = \beta D^*(t) l_0 \frac{\partial \bar{C}_\theta}{\partial x},
\end{equation}
where the gradient is across the mixing front and $\beta$ is a proportionality constant that includes the original diffusion/dispersion coefficient and the dimensionless quantities defined in Eq.~\ref{eq:dimensionlessQuantities}. Note that ${\partial \bar{C}_\theta}/{\partial x} \sim O\left({\Delta C_\theta}/{\delta_0}\right)$, where $\Delta C_\theta$ is the change in concentration across the front (here $\Delta C_\theta = 1$). 

The diffusive mass transfer in the 2D problem, $J_{2D}$, is given by
\begin{equation} \label{eq:2Dflux}
    J_{2D} = \beta \int_{\Gamma}\frac{\partial C_\theta}{\partial n} d\Gamma, \ \ {\text{where}} \ \ \frac{\partial C_\theta}{\partial n} = \nabla C_\theta \cdot \mathbf{n},
\end{equation}
with $\mathbf{n}$ the normal to $\Gamma$ and $\beta$ the same constant as in Eq.~\ref{eq:1Dflux}. This can be rewritten in terms of the length of the mixing front and the average concentration gradient across the interface
\begin{equation} \label{eq:2Dflux_avgGradient}
    J_{2D} = \beta l(t) \left\langle \frac{\partial C_\theta}{\partial n} \right\rangle.
\end{equation}
Here we use the $\langle \cdot \rangle$ notation (rather than an overbar) to emphasize that the full gradient is averaged. The average gradient is computed along the mixing front and can be expressed as
\begin{equation} \label{eq:ave_grad}
    \left\langle \frac{\partial C_\theta}{\partial n} \right\rangle = \frac{1}{l(t)} \int_{\Gamma}\frac{\partial C_\theta}{\partial n} d\Gamma.
\end{equation}
Importantly, ${\partial C_\theta}/{\partial n}$ and $\langle {\partial C_\theta}/{\partial n} \rangle$ are both $O\left({\Delta C_\theta}/{\bar{\delta}(t)} \right)$, which is a factor of ${\delta_0}/{\bar{\delta}(t)}$ larger than ${\partial \bar{C}_\theta}/{\partial x}$, the gradient in the 1D problem. With this we have
\begin{equation} \label{eq:approxConcGradients}
    \left\langle \frac{\partial C_\theta}{\partial n} \right\rangle \approx \frac{\delta_0}{\bar{\delta}(t)} \frac{\partial \bar{C}_\theta}{\partial x}.
\end{equation}

There are two early-time effects in the fully resolved 2D problem -- the stretching of the mixing front and the associated reduction in thickness, leading to larger concentration gradients across the interface \citep{ranz1979applications} -- that are not captured explicitly in the 1D representation. These effects must, therefore, be incorporated into $D^*(t)$. By equating $J_{1D}$ in Eq.~\ref{eq:1Dflux} to $J_{2D}$ in Eq.~\ref{eq:2Dflux_avgGradient}, and using Eq.~\ref{eq:approxConcGradients}, we arrive at 
\begin{equation} \label{eq:Dstar_noDiff}
    D^*_{et}(t) \approx \frac{l(t)} {\bar{\delta}(t)}\frac{\delta_0}{l_0} = l^2(t),
\end{equation}
where the subscript $et$ denotes `early-time.' This 1D (upscaled) representation is valid only for times where the diffusive smearing of the mixing front is small. Note that the appearance of a coefficient $\xi$, such that $D^*_{et}(t) = \xi \times l^2(t)$ in Eq.~\ref{eq:Dstar_noDiff}, is also consistent with the scaling arguments provided above. As we will see, we achieve an accurate approximation using $\xi=1$, which suggests that the variations in $\delta$ along the 2D front have little effect on $\langle {\partial C_\theta}/{\partial n} \rangle$.

At later times, diffusive effects will counteract the stretching introduced by variations in the velocity field, and $D^*(t)$ will gradually approach an asymptotic value, $\Dsm$. This quantity is analogous to the macroscopic diffusion coefficient in Taylor dispersion theory. In order to represent the early- and late-time behaviors, along with an intermediate regime that links them, we propose the following general form for $D^*(t)$
\begin{equation} \label{eq:Dstar}
    D^*(t) = \frac{D^*_{max} l^2(t)} {D^*_{max} + l^2(t) - 1}.
\end{equation}
Here ${D}^*_{max}$ is the only free parameter in the upscaled representation. Note that Eq.~\ref{eq:Dstar} has the proper limiting behaviors, as it gives $D^*(t=0) = 1$, $D^*(t) \approx l^2(t)$ at early-times, and $D^*(t \rightarrow \infty) = D^*_{max}$.

We reiterate that no explicit assumptions regarding the mixing front geometry are required in the derivation of Eq.~\ref{eq:Dstar_noDiff}. We thus expect this representation to be valid for deformation patterns that differ from that in Figure~\ref{fig:stretch}. Indeed, even though the methodology is presented and applied here to upscale 2D problems to 1D, we expect it to also be applicable for 3D problems. In this case, the mixing front length, $l(t)$, would be replaced by a mixing front area. It is also important to note that functional forms other than Eq.~\ref{eq:Dstar} could be used to represent (and interpolate between) the short- and long-time behaviors we seek to capture. Although we do not claim that the precise form in Eq.~\ref{eq:Dstar} is optimal, the results presented in Section~\ref{sec:applications} demonstrate that it provides high levels of accuracy for the cases considered.

\subsubsection*{Computation of $l(t)$}
We use particle tracking to construct $l(t)$ from the variable velocity field that appears in the 2D (fine-scale) problem. Our specific approach entails launching a large number of equally spaced particles along the interface $\Gamma$, and then tracking the evolution of their positions in time. A forward Euler approximation is used for this purpose
\begin{equation}
    \mathbf{x}(t + \Delta t) = \mathbf{x}(t) + \mathbf{v}\left(\mathbf{x}(t)\right) \Delta t,
\end{equation}
where $\mathbf{x}$ denotes particle location and $\Delta t$ is the time step size, which is chosen to be the same as the time step used in the spectral solution (discussed in the appendix). At any time $t$, $l(t)$ can be directly computed by summing the distances between the positions of the particles. With this approach, $l(t)$ is obtained at very little computational cost.

\subsubsection*{Estimates of $\Dsm$} \label{sec:predictiveApproach}

We now apply scaling procedures to relate $D^*_{max}$ to system variables. For the particular case of the oscillatory velocity field, we can obtain an exact analytical expression for $l(t)$. This expression is very well approximated by
\begin{equation} \label{eq:approx_l2}
    l^2(t) \approx 1 + \left(4 a \omega \lambda t \right)^2.
\end{equation}
Although we apply particle tracking to construct $l(t)$ in all of our results, Eq.~\ref{eq:approx_l2} is useful for the scaling arguments introduced below.

The system will reach a Taylor-dispersion-like limit when the diffusion length is of $O(\tilde{L}_d)$, where $\tilde{L}_d$ is the length between extrema in the front, shown in Figure~\ref{fig:stretch}c. The characteristic diffusion time ($\ttm_d$) for this to occur is $\ttm_d \propto \tilde{L}_d^2/\tilde{D}_0$. To express this relation in dimensionless form, we introduce $L_d = \tilde{L}_d / \Lyt$ and $t_d = \ttm_d/\ttm_c$. After recognizing that, for the oscillatory velocity field $L_d = 1/\omega$, we obtain

\begin{equation} \label{eq:td_adim}
    t_d \propto \frac{Pe}{\omega^2} \frac{1}{\lambda}.
\end{equation}

It is useful to introduce a quantity $\alpha$, defined as
\begin{equation} \label{eq:alpha}
    \alpha = \frac{D^*(t_{na})-1}{D^*_{max}-1},
\end{equation}
where $t_{na}$ is the time required to achieve a specified fraction of the increase in $D^*(t)$ (from 1 to $D^*_{max}$). The subscript $na$ denotes near-asymptotic. The parameter $\alpha$ is user specified, though here we are interested in system behavior for $\alpha$ near 1. If we set $\alpha=0.9$, this means that, at a time $t_{na}$, $D^*(t)$ will have reached a value of $1+0.9(D^*_{max}-1)$. We expect $t_{na}$ (for $\alpha$ near 1) to be directly related to $t_d$ in Eq.~\ref{eq:td_adim}, such that we can write $t_{na}=\kappa t_d$, where $\kappa$ is a constant.

Our goal at this point is to determine how $D^*_{max}$ scales with system parameters. Introducing the expression for $D^*(t)$ given in Eq.~\ref{eq:Dstar} into Eq.~\ref{eq:alpha}, and rearranging to provide an expression for $D^*_{max}$, we have
\begin{equation} \label{eq:DstarMax_general}
    D^*_{max} = \frac{l^2(t_{na})\left[1-\alpha\right]+\alpha-1}{\alpha}.
\end{equation}
Substituting Eq.~\ref{eq:approx_l2} in Eq.~\ref{eq:DstarMax_general} gives

\begin{equation} \label{eq:DstarMax_sines_t}
    D^*_{max} \approx \frac{1-\alpha}{\alpha} \left(4a \omega \lambda t_{na}\right)^2.
\end{equation}
Now, using Eq.~\ref{eq:td_adim} combined with the fact that $t_{na} = \kappa t_d$, we can rewrite Eq.~\ref{eq:DstarMax_sines_t} as

\begin{equation} \label{eq:DstarMax_intermed}
    D^*_{max} \propto \frac{16\kappa^2(1-\alpha)}{\alpha} \Psi,
\end{equation}
with
\begin{equation} \label{eq:Psi}
    \Psi = Pe^2 a^2/\omega^2.
\end{equation}
Importantly, Eq.~\ref{eq:DstarMax_intermed} shows that $\Dsm$ is proportional to $\Psi$. Considering that $\Dsm$ should be 1 at the limits $Pe \rightarrow 0$, $a \rightarrow 0$, and $\omega \rightarrow \infty$, we write

\begin{equation} \label{eq:DstarMax_fit}
    D^*_{max} \approx 1 + c \Psi.
\end{equation}
Here $c$ is a constant, which can be determined from reference results for a few cases, as will be explained in the next section.

\subsection{Upscaling workflow for reactive transport problems} \label{sec:workflow}

Consistent with our earlier discussion, we propose an upscaled (1D) form of Eq.~\ref{eq:conserv_adim} that relies on the upscaled model for mixing presented in Section~\ref{sec:upscaledMixing}. The effects of the variable velocity field and consequent front deformation on the reactive transport problem are captured by $D^*(t)$. The dimensionless equations are 

\begin{subequations}\label{eq:conserv_adim_upsc}
    \begin{align}
        \frac{\partial \bar{C}_i}{\partial t} +  \frac{\partial \bar{C}_i}{\partial x}= \frac{D^*(t)}{Pe}  \frac{\partial^2 \bar{C}_i}{\partial x^{2}} - Da \bar{C}_A \bar{C}_B , \quad 0 < x < \lambda, \quad i = \textit{A} \text{ and } \textit{B},\\
        \frac{\partial \bar{C}_C}{\partial t} +  \frac{\partial \bar{C}_C}{\partial x}= \frac{D^*(t)}{Pe}  \frac{\partial^2 \bar{C}_C}{\partial x^{2}} + Da \bar{C}_A \bar{C}_B , \quad 0 < x < \lambda.
    \end{align}
\end{subequations}

Our goal is to apply this 1D representation for a large number of cases without having to first solve the full set of 2D equations for each case. In order to accurately predict $\Dsm$ (required in Eq.~\eqref{eq:Dstar}), however, we must first determine the value of $c$ in Eq.~\ref{eq:DstarMax_fit}. To accomplish this, we do need to solve the full 2D system for a few select `training' cases. These cases should correspond to a wide range of $\Psi$ values. 

For these training cases, we first solve Eq.~\ref{eq:conserv_adim} to obtain reference results. This is the most time-consuming step of the overall procedure. Given a 2D solution, we then solve an optimization problem to determine the value of $D^*_{max}$ in Eq.~\ref{eq:conserv_adim_upsc} such that a relevant metric quantifying the difference between the 2D and 1D problems is minimized.

The metric used in this work is the conversion factor of the chemical reaction, denoted by $\eta$. This is defined for the 2D and 1D problems as

\begin{subequations}\label{eq:convFactors}
    \begin{align}
        \eta_{2D}(t) = \frac{1}{m_c^{max}} \int_{0}^{1}\int_{0}^{\lambda} C_C(x,y,t)dxdy,\\
        \eta_{1D}(t) = \frac{1}{m_c^{max}} \int_{0}^{\lambda} \bar{C}_C(x,t)dx,
    \end{align}
\end{subequations}
where $m_c^{max}$ is the maximum amount of component $C$ that would be formed under 100\% conversion of the reactants. The parameter $\Dsm$ is now obtained by solving the optimization problem

\begin{equation}\label{eq:invProblem}
   \Dsmopt =  \argmin_{\Dsm} \left< \left[\eta_{2D}(t) - \eta_{1D}(t)\right]^2 \right>,
\end{equation}
where $\langle \cdot \rangle$ represents the time average. Thus, we find $\Dsm$ such that the time average of the difference between the 1D and 2D conversion factors is minimized. Results from the 1D model are linearly interpolated in time to provide solutions at the same times as the 2D model. Although $\Dsm$ is determined through consideration of the conversion factors, the resulting 1D solutions also provide reasonable representations of the spatial distribution of the chemical reactions, as will be shown in Section~\ref{sec:applications}. The single-variable minimization problem in Eq.~\ref{eq:invProblem} is solved using a truncated Newton method implemented in Python's \textit{scipy} library \citep{virtanen2020scipy}. 

Use of this $D^*_{max}$ in Eq.~\ref{eq:Dstar} provides the full $D^*(t)$ needed for the upscaled representation in Eq.~\ref{eq:conserv_adim_upsc}. However, as noted earlier, $D^*_{max}$ is only obtained in this way for a few (e.g., $\sim$4) training cases. Given the training case solutions, the coefficient $c$ in Eq.~\ref{eq:DstarMax_fit} is determined through a linear fit. Then, for the rest of the parameter values considered, $D^*_{max}$ is assigned using Eq.~\ref{eq:DstarMax_fit}. Thus, the need for 2D reference solutions is avoided for the large majority of cases.

To reiterate, the general workflow consists of the following steps:
\begin{itemize}
    \item Step 1 -- Select a few training cases (i.e., values of the parameters $Pe$, $a$ and $\omega$) that span a wide range of $\Psi$ values. 
    \item Step 2 -- Solve Eq.~\ref{eq:conserv_adim} (the full 2D system) for each training case.
    \item Step 3 -- Solve the optimization problem, Eq.~\ref{eq:invProblem}, to obtain $\Dsmopt$ values for each of the training cases.
    \item Step 4 -- Apply a linear fit to determine $c$ in Eq.~\ref{eq:DstarMax_fit} from the training-case $\Dsm$ values.
    \item Step 5 -- For all new parameter sets of interest, obtain $\Dsm$ with Eq.~\ref{eq:DstarMax_fit} using the calibrated value of $c$. Then solve only the upscaled 1D model given by Eq.~\ref{eq:conserv_adim_upsc}. 
\end{itemize}

We note finally that the upscaled model described in this section can also be applied to quantify mixing in non-reactive transport problems. In this case, the general steps presented above still apply, though the conversion factors appearing in Eq.~\ref{eq:invProblem} are replaced by an appropriate mixing metric. The use of the upscaled model in this setting will be considered in Section~\ref{sec:oscillatoryNoReac}. 

\section{Results using upscaled model} \label{sec:applications}

We now apply our upscaling approach to two distinct types of flow fields -- the oscillatory velocity field and the ABC field. Based on numerical convergence tests shown in the appendix, we use 256 grid points in each coordinate direction in the spectral solutions. This means the 2D models contain 256$\times$256 = 65,536 points, while the upscaled models involve only 256 points. The final time is $t_f = 10$, i.e., 10 revolutions around the periodic domain. Additionally, 50,000 particles are used in the particle-tracking computation of $l(t)$. The computational cost of this is negligible compared to even the solution of the 1D models.

In each scenario we consider a large set of runs. To avoid periodicity effects, we postprocess the runs and remove cases where the concentration of a given component interacts with itself via periodicity. Cases with $\eta(t_f) > 0.9$ are also excluded, as this condition represents a near depletion of the reactants. Numerical experiments have shown that the solution of Eq.~\ref{eq:invProblem} is somewhat sensitive to the final time of the simulations ($t_f$). This is as expected since, if the simulation time is too short, the near-asymptotic behavior has not been reached and a reasonable estimate of $\Dsmopt$ cannot be obtained. For this reason, we also eliminate cases for which near-asymptotic behavior has not been reached, namely cases where $\alpha(t_f) > \alpha_{min}$. The quantity $\alpha_{min}$ is a number close to 1 and can vary from case to case (values for $\alpha_{min}$ will be given below).

The average and maximum errors in the upscaled model in the representation of $\eta$ are as follows:

\begin{subequations}\label{eq:errors}
    \begin{align}
        e_{avg} = \left< \frac{|\eta_{2D}(t) - \eta_{1D}(t)|}{\eta_{2D}(t)} \right> \\
        e_{max} = \max \left(\frac{|\eta_{2D}(t) - \eta_{1D}(t)|}{\eta_{2D}(t)}\right)
    \end{align}
\end{subequations}
We are also interested in analyzing the accuracy of the upscaled model in the representation of the spatial distribution of the chemical reaction. For this, we solve a conservation equation for a component $C_s$, where the subscript $s$ indicates solid. This equation is equivalent to the conservation equation for component $C$, but without the transport terms. The deposition of $C_s$ does not affect the flow properties. Visualization of $C_s$ shows the locations in the domain where the reaction occurs, with large $C_s$ meaning strong reaction intensity.

\subsection{Oscillatory velocity field} \label{sec:oscillatory}

We first consider velocity fields given by Eq.~\ref{eq:velFieldOsc}. A total of 72 cases, spanning all possible combinations attainable with $\omega = (4, 8, 12, 16, 20, 24)$, $Pe = (1000, 2000, 5000, 10,000, 20,000, 40,000)$, and $a = (0.05, 0.1)$, are simulated. With $\alpha_{min}=0.9$, 43 of the 72 cases pass the filters described above. Most of the excluded runs are those for which $\omega$ is small and $Pe$ is large, as these runs tend to either not reach $\alpha(t_f) > \alpha_{min}$, or they are affected by periodicity.

\begin{figure} [H]
    \centering
    \includegraphics[width=0.49\textwidth,trim = 220 0 100 0,clip]{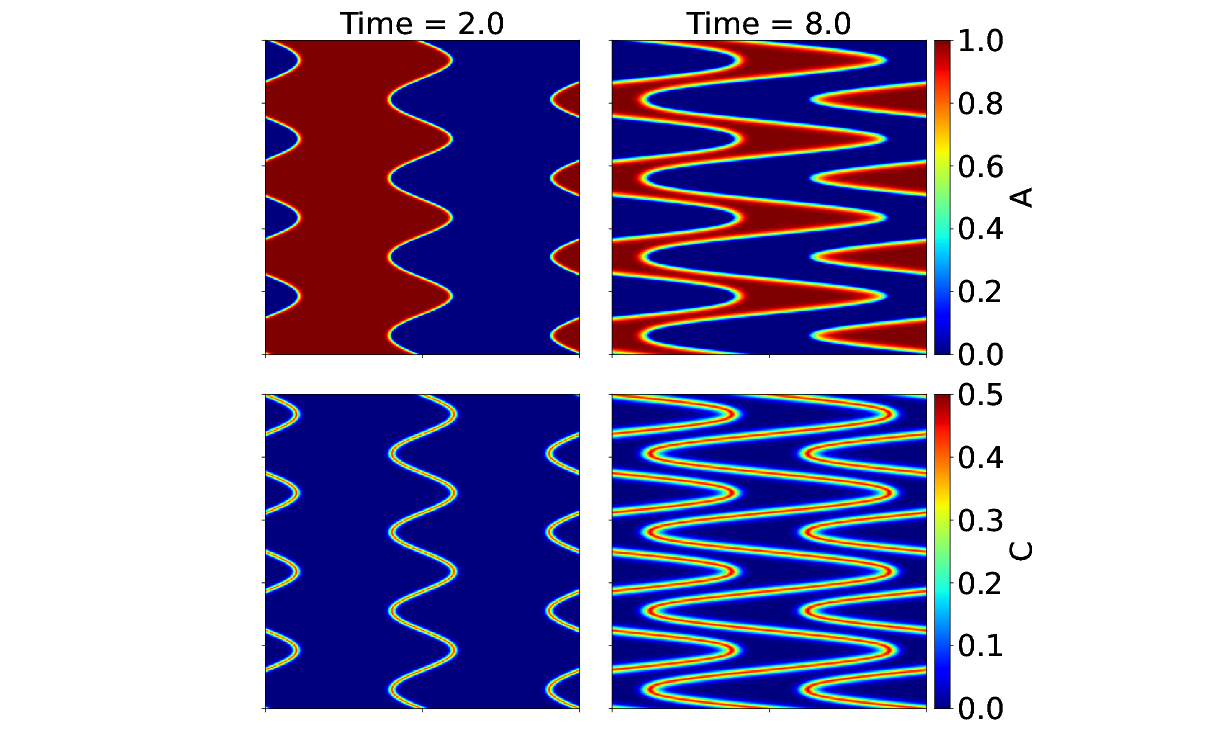}
    \caption{Concentration of component $A$ (top row) and $C$ (bottom row) at $t = 2$ and $t = 8$ for $Pe = 40,000$, $\omega= 4$ and $a = 0.05$.}
    \label{fig:maps2D_Pe40000_w4_a0.5_Oscil}
\end{figure}

To illustrate some general behaviors, we present results for two cases -- one that reaches near-asymptotic behavior slowly, and one that does so quickly (according to Eq.~\ref{eq:td_adim}). Figure~\ref{fig:maps2D_Pe40000_w4_a0.5_Oscil} displays the concentration of $A$ and $C$ at $t=2$ and $t=8$ for a case with $Pe=40,000$ and $\omega=4$. We see that, from $t = 2$ to $t=8$, there is a significant increase in the length of the reactive front. Importantly, however, the interface has not yet `coalesced' through the effects of diffusion. This demonstrates that this system requires a long time to reach near-asymptotic behavior. We will see later that $D^*(t)$ has not yet plateaued for this case.

\begin{figure} [H]
    \centering
    \includegraphics[width=0.49\textwidth,trim = 220 0 100 0,clip]{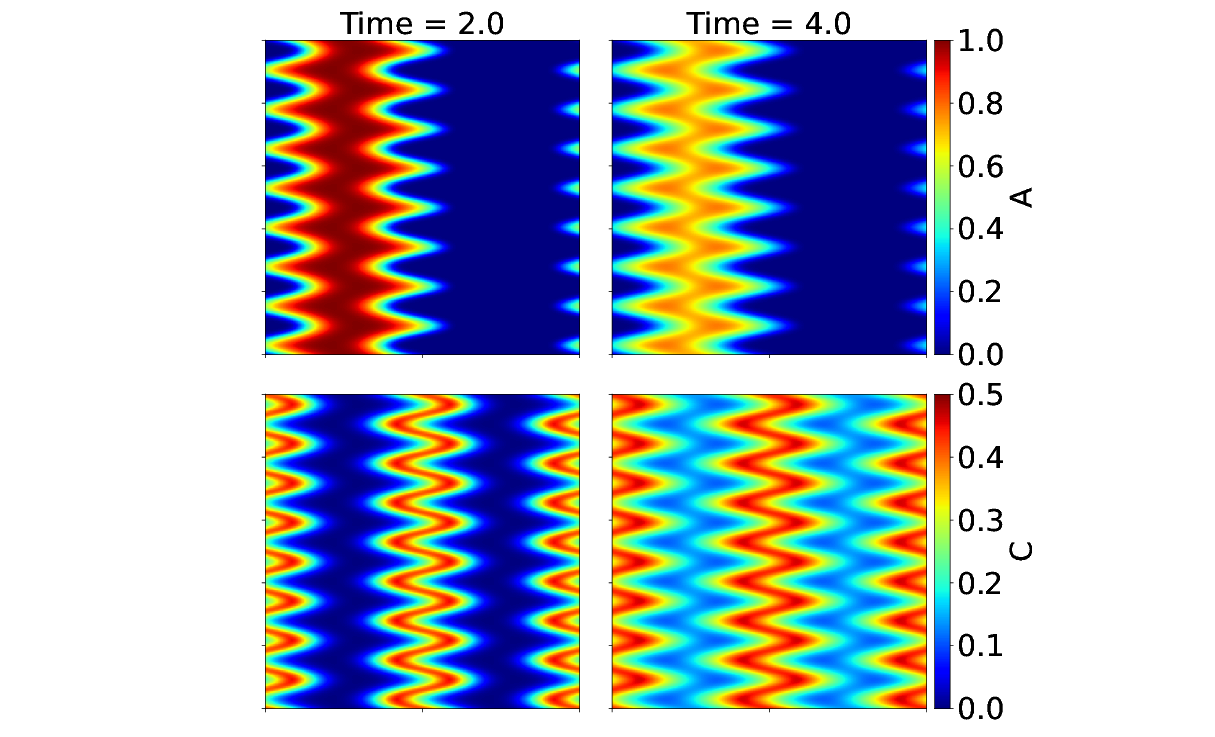}
    \caption{Concentration of component $A$ (top row) and $C$ (bottom row) at $t = 2$ and $t = 4$ for $Pe = 2000$, $\omega = 8$ and $a = 0.1$.}
    \label{fig:maps2D_Pe2000_w8_a0.1_Oscil}
\end{figure}

Different behavior is evident in Figure~\ref{fig:maps2D_Pe2000_w8_a0.1_Oscil}, where we show results for $Pe = 2000$ and $\omega = 8$. Here, because there are stronger diffusive effects and less distance between extrema in the front profile ($L_d$), the system reaches near-asymptotic behavior much faster. From $t=2$ to $t=4$, there is no marked increase in the length of the reactive front. Thus, even though $l(t)$ continues to increase (as it is a purely advective quantity), the concentration profile for component $A$ becomes smoother. Consistent with this behavior, $D^*(t)$ reaches its plateau value at around $t=4$, as will be seen below.

\begin{figure} [H]
    \centering
    \includegraphics[width=0.49\textwidth,trim = 0 0 0 0,clip]{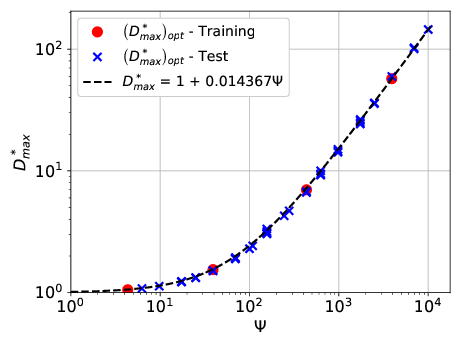}
    \caption{$\Dsm$ results for the oscillatory velocity field. The blue $\times$'s and red circles represent values of $\Dsmopt$ computed from Eq.~\ref{eq:invProblem}. The four red circles are the training (calibration) points for the fit, shown by the dashed curve.}
    \label{fig:trainingOscillatoryReactive}
\end{figure}

We now focus on the prediction of reaction behavior from upscaled models. This first requires the determination of the fitting constant $c$ in Eq.~\ref{eq:DstarMax_fit}. To accomplish this, four training runs, corresponding to the $\Psi$ values closest to 5, 50, 500, and 5000, are selected. Figure~\ref{fig:trainingOscillatoryReactive} displays a log plot of $\Dsm$ versus $\Psi$ for all 43 runs. The values for all 43 points are the $\Dsmopt$ computed from Eq.~\ref{eq:invProblem}, i.e., from fitting 1D solutions to the 2D results. The four training runs, shown as the red points, are used to construct the fit. It is evident that the fit is highly accurate, confirming the validity of the scaling arguments presented in Section~\ref{sec:predictiveApproach}.

Using the calibrated value of $c = 0.014367$, we construct upscaled results for all 43 cases. The largest average error ($e_{avg}$ in Eq.~\ref{eq:errors}) over the 39 test (predictive) cases is 3.29\%, and the largest maximum error ($e_{max}$) is 7.35\%. Summary results are presented in Table~\ref{tableOscil}. The table provides the input parameters as well as $e_{avg}$ and $e_{max}$ for the four training cases and for five predictive cases. The latter are labeled (a)-(e) and will be referred to later. Four of the predictive cases pass the filters, while case~(e) does not satisfy $\alpha(t_f) > 0.9$. The five predictive cases correspond to (a) smallest $\eta$, (b) an additional case with $a = 0.1$, (c) largest $\eta$, (d) largest $e_{max}$, and (e) a case with small $\alpha$ where the early-time approximation of Eq.~\ref{eq:Dstar_noDiff} is tested with large stretching effects.

\begin{table}[H]
\caption{\label{tableOscil} Input parameters and errors for training and predictive (test) runs with oscillatory velocity field. Predictive cases are labeled for later reference.}
\centering
\begin{tabular}{ccccccc}
\hline
Type           & $Pe$ & $a$  & $\omega$ & $\Psi$ & $e_{avg}$ & $e_{max}$ \\ \hline
Training       & 1000  & 0.05 & 24       & 4.3    & 1.51E-03  & 1.19E-02  \\ 
Training       & 1000  & 0.05 & 8        & 39.1   & 8.85E-03  & 2.59E-02  \\
Training       & 5000  & 0.05 & 12       & 434.0  & 7.25E-03  & 1.92E-02  \\
Training       & 20,000  & 0.05 & 16       & 3906.3 & 2.11E-02  & 6.74E-02  \\
Predictive (a) & 5000  & 0.05 & 24       & 108.5  & 2.04E-02  & 3.04E-02  \\
Predictive (b) & 2000  & 0.10 & 8        & 625.0  & 5.61E-03  & 3.38E-02  \\
Predictive (c) & 10,000   & 0.10 & 16       & 3906.3 & 1.94E-02  & 6.70E-02  \\
Predictive (d) & 40,000  & 0.05 & 20       & 10,000  & 2.92E-02  & 7.35E-02  \\
Predictive (e) & 40,000  & 0.05 & 4        & 250,000 & 1.31E-02  & 3.15E-02 \\ \hline
\end{tabular}
\end{table}

\begin{figure} [H]
     \centering
     \begin{subfigure}[b]{0.49\textwidth}
         \centering
         \includegraphics[width=\textwidth]{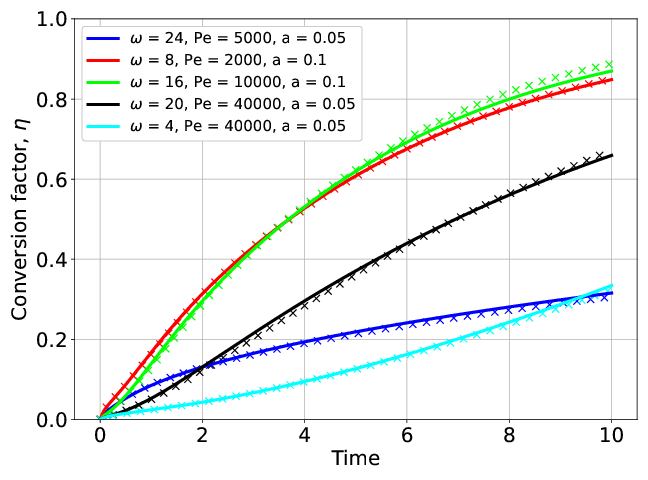}
         \caption{}
     \end{subfigure}
     \hfill
     \begin{subfigure}[b]{0.49\textwidth}
         \centering
         \includegraphics[width=\textwidth]{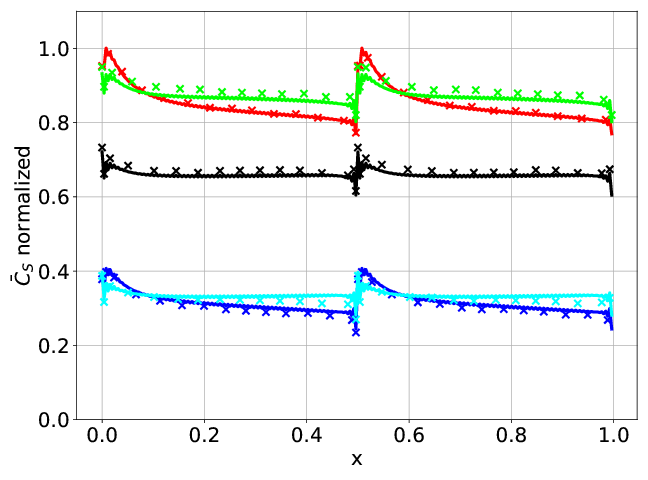}
         \caption{}
     \end{subfigure}
     \hfill
     \begin{subfigure}[b]{0.49\textwidth}
         \centering
         \includegraphics[width=\textwidth]{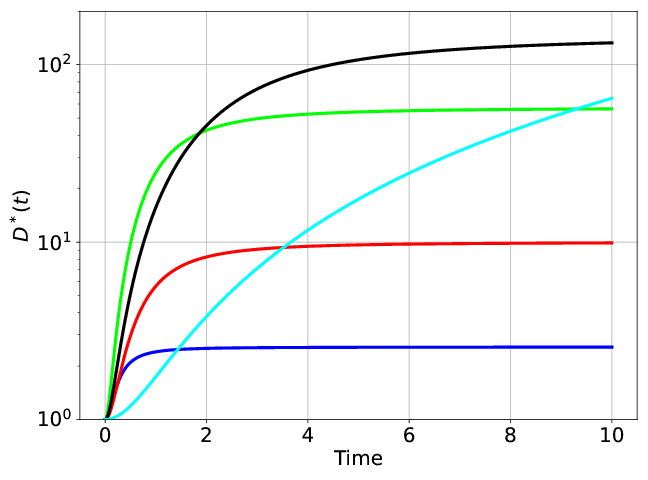}
         \caption{}
     \end{subfigure}
             \caption{(a) Conversion factor for the five predictive cases in Table~\ref{tableOscil}; (b) normalized $\bar{C}_s$ at $t=10$; (c) $D^*(t)$ in the upscaled representation. The $\times$'s depict results from the reference 2D models and the solid curves show the upscaled 1D representation. Legend in (a) also applies to (b) and (c).}
        \label{fig:convFactor_Dstar_oscil}
\end{figure}

Figure~\ref{fig:convFactor_Dstar_oscil}a displays the evolution of conversion factor with time for the five predictive runs in Table~\ref{tableOscil}. In this and all subsequent figures, the solid curves depict the 1D upscaled results, while the $\times$'s display the 2D (fine-scale) reference results. We emphasize that these 1D results are constructed independent of their corresponding 2D reference results; i.e., they use $\Dsm$ computed from the fit (Eq.~\ref{eq:DstarMax_fit}), not $\Dsmopt$ from fine-scale 2D simulation. Excellent agreement between the 2D and 1D representations, over a wide range of parameter values and conversion factor behavior, is clearly observed. 

Figure~\ref{fig:convFactor_Dstar_oscil}b shows results for $C_s$ at $t=10$. The 2D results are averaged in the $y$-direction, and both sets of results are normalized by the amount of $C_s$ that would be formed by uniformly reacting all of components $A$ and $B$ initially present throughout the domain. In Figure~\ref{fig:convFactor_Dstar_oscil}b we see that, even though $\Dsm$ is computed to match the time evolution of the conversion factor, the upscaled model also provides a reasonable representation of the spatial distribution of the chemical reaction. Figure~\ref{fig:convFactor_Dstar_oscil}c shows $D^*(t)$ in the upscaled representation. Notably, in one case $D^*(t)$ reaches a value greater than 100, indicating that the model is able to incorporate a very substantial amount of stretching-enhanced mixing into the upscaled dispersion coefficient.

Also of interest is the observation that $D^*(t)$ for predictive case~(e) ($\Psi = 250,000$) closely follows the early-time approximation given in Eq.~\ref{eq:Dstar_noDiff} over the entire time frame. The agreement between the fine and upscaled representations seen in the cyan curves in Figure~\ref{fig:convFactor_Dstar_oscil}a and b for this case illustrates the accuracy of the upscaled model under conditions of significant stretching of the mixing front without appreciable diffusive coalescence. It is also important to observe that, although the four training cases consider $a = 0.05$, accurate predictions are nonetheless obtained with $a = 0.1$. Additionally, the maximum $Pe$ and $\Psi$ in the training runs are 20,000 and 3906, while accurate results are achieved for $Pe = 40,000$ and $\Psi = 250,000$. These observations further demonstrate the appropriateness of our scaling arguments and treatments.

In summary, the results in Figure~\ref{fig:convFactor_Dstar_oscil} show that our upscaled model captures the most important physical phenomena controlling the mixing and reactive transport processes, even in cases where large amounts of reactive front stretching and deformation are experienced. We note, however, that this upscaled representation will not in general provide an accurate description of the spreading of the plumes of components $A$, $B$, and $C$. The proper description of spreading would require an apparent dispersion coefficient that captures second moments of the concentrations. As discussed in the Introduction, a dispersion of this type would lead to the overestimation of the reaction rate. In other words, the upscaled dispersion coefficient can describe mixing or macrodispersion, but not both \citep{Cirpka2002}. In future work, it may be of interest to develop a two-equation representation that is able to capture both of these effects.

\subsubsection*{Mixing of a conservative component} \label{sec:oscillatoryNoReac}

We now demonstrate that the upscaled model can be used to model the mixing of a conservative (non-reactive) component $A$. The setup is the same as was considered above, except here we set $Da = 0$. As mentioned in Section~\ref{sec:workflow}, for the non-reactive case the conversion factor $\eta$ (in Eq.~\ref{eq:invProblem}) is replaced by an appropriate mixing metric. Here we take this mixing metric to be the concentration variance of $A$ \citep{cushman2016handbook}, denoted $\sigma^2_A$,

\begin{equation} \label{eq:covariance}
    \sigma^2_A(t) = \int_{\Omega} \left\langle \left(C_A(\mathbf{x},t) - \bar{C}_A\right)^2\right\rangle d\Omega,    
\end{equation}
where $\int_\Omega$ represents integration over the full domain, and the average concentration is constant ($\bar{C}_A = 0.5$) as there is no chemical reaction. Note that in the minimization problem (Eq.~\ref{eq:invProblem}) we use the relative error in this case instead of the absolute error.

\begin{figure} [H]
     \centering
     \begin{subfigure}[b]{0.49\textwidth}
         \centering
         \includegraphics[width=\textwidth]{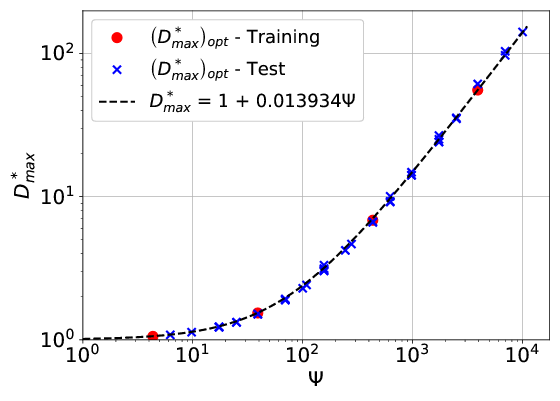}
         \caption{}
     \end{subfigure}
     \hfill
     \begin{subfigure}[b]{0.49\textwidth}
         \centering
         \includegraphics[width=\textwidth]{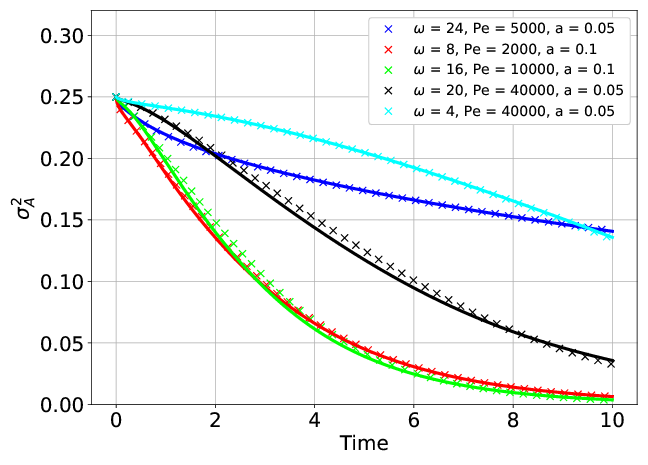}
         \caption{}
     \end{subfigure}
         \caption{(a) $\Dsm$ results for a conservative component for the oscillatory velocity field. The blue $\times$'s and red circles represent values of $\Dsmopt$ computed via optimization. The four red circles are the training points for the fit, shown by the dashed curve. (b) Concentration variance, $\sigma^2_A$. The $\times$'s depict results from the reference 2D models and the solid curves show the upscaled 1D representation.}
        \label{fig:trainingAndCovarianceNonReac}
\end{figure}

We again predict $\Dsm$ from a fit (Eq.~\ref{eq:DstarMax_fit}) involving the same four training runs.  Figure~\ref{fig:trainingAndCovarianceNonReac}a shows $\Dsm$ for these training runs and for the 39 test runs (again, these are the same as in Figure~\ref{fig:trainingOscillatoryReactive}). The values of $\Dsm$ computed for the non-reactive case are very similar to those for the reactive case. Specifically, for the reactive case we found $c=0.014367$, and for the non-reactive case $c=0.013934$ -- a difference of only about 3\%. This suggests that the parameter $c$ can be determined at less computational cost from results for a single conservative component instead of from multicomponent reactive transport computations. This is similar to what was done by \cite{willmann2010coupling} to calibrate unknown parameters in a multirate mass transfer model for mixing-limited reactions. Figure~\ref{fig:trainingAndCovarianceNonReac}b displays the time evolution of $\sigma^2_A$ for the five predictive cases defined in Table~\ref{tableOscil}. Excellent accuracy in $\sigma^2_A$ is observed, confirming the applicability of our upscaled mixing model to a related problem involving non-reactive transport.

\subsection{ABC velocity field}

We now assess the performance of the upscaled model for the velocity field given by Eq.~\ref{eq:velFieldABC}, with $N_f = 1$, $\bar{v}_y = 0$, and $\gamma = 0$. Although $\bar{v}_y = 0$, $v_y$ is locally nonzero. A total of 60 cases that span all possible combinations attainable with $\omega = (2, 4, 6, 8, 10, 12)$, $Pe = (1000, 2000, 5000, 10000, 20000)$, and $a = (0.35, 0.45)$ are considered. Of these 60 cases, 41 pass the filters described earlier, with $\alpha_{min}=0.915$.

\begin{figure} [H]
    \centering
    \includegraphics[width=0.99\textwidth,trim = 100 0 70 0,clip]{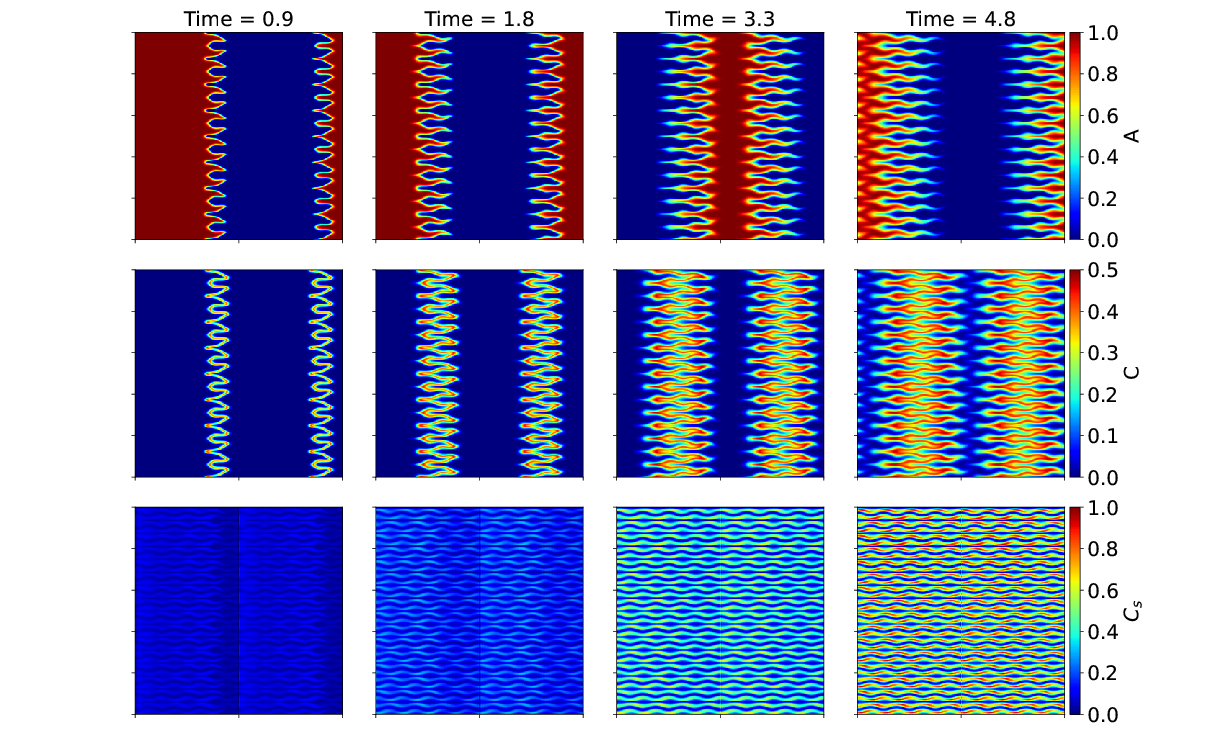}
    \caption{Concentration of component $A$ (top row), component $C$ (middle row) and normalized $C_s$ (bottom row) at $t$ = 0.9, 1.8, 3.3 and 4.8 for $Pe = 20,000$, $\omega = 8$ and $a = 0.35$ for ABC velocity field.}
    \label{fig:maps2D_Pe20000_w8_a0.35_ABC_v1}
\end{figure}

The 2D nature of the ABC velocity field leads to complex mixing front geometry, as well as an intricate spatial distribution of reaction locations ($C_s$). Results for a case with $Pe=20,000$, $\omega=8$ and $a=0.35$ are shown in Figure~\ref{fig:maps2D_Pe20000_w8_a0.35_ABC_v1}. There we see substantial spreading of the front due to the complex velocity field (top and middle rows) and
the presence of reaction `hot spots' along with locations where very little reaction occurs (bottom row). We will show that our upscaling approach can provide reasonable 1D representations of the complex behaviors observed in Figure~\ref{fig:maps2D_Pe20000_w8_a0.35_ABC_v1}.

\begin{figure} [H]
    \centering
    \includegraphics[width=0.49\textwidth,trim = 0 0 0 0,clip]{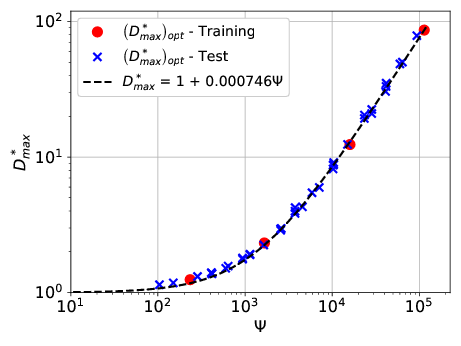}
    \caption{$\Dsm$ results for the ABC velocity field. The blue $\times$'s and red circles represent values of $\Dsmopt$ computed from Eq.~\ref{eq:invProblem}. The four red circles are the training (calibration) points for the fit, shown by the dashed curve.}
    \label{fig:trainingABCReactive}
\end{figure}

Equation~\ref{eq:DstarMax_fit}, with $\Psi = Pe^2 a^2 / \omega^2$, was derived through consideration of the oscillatory velocity field for which $l^2(t) \propto w^2a^2t^2$. For the ABC velocity field, $l^2(t) \propto w^2a^4t^2$, indicating the appropriate definition of $\Psi$ is $\Psi = Pe^2 a^4/\omega^2$. To calibrate $c$ in Eq.~\ref{eq:DstarMax_fit}, four training runs are selected. Two of these cases have $a = 0.35$ and two have $a = 0.45$; otherwise we select cases  that have values of $\Psi$ closest to 200, 2000, 20,000, and 200,000. Figure~\ref{fig:trainingABCReactive} displays the relationship between $\Dsm$ and $\Psi$ for the 41 runs considered. The points are defined as in Figure~\ref{fig:trainingOscillatoryReactive}. As was the case for the oscillatory velocity field, excellent agreement between $\Dsm$ obtained with the fit (Eq.~\ref{eq:DstarMax_fit}) and by solving the minimization problem  (Eq.~\ref{eq:invProblem}) is observed. This is encouraging as it reinforces the scaling arguments underlying the model, and it demonstrates that the definition of $\Psi$ ($Pe^2 a^4/\omega^2$) is appropriate for this case.

Using the calibrated value of $c$, upscaled results are obtained for all 41 cases. The largest $e_{avg}$ observed over all cases is 4.10\%, and the largest $e_{max}$ is 6.12\% (these errors are as defined in Eq.~\ref{eq:errors}). Input parameters and errors for the four training cases and for five predictive cases are shown in Table~\ref{tableABC}. The five predictive cases correspond to (a) largest mean error, (b) smallest $\eta$, (c) an additional case with small $\omega$ and intermediate value of $\Dsm$, (d) largest $\eta$, and (e) largest $\Psi$.

\begin{table}[H]
\caption{\label{tableABC} Input parameters and errors for training and predictive (test) runs with ABC velocity field. Predictive cases are labeled for reference.}
\centering
\begin{tabular}{ccccccc}
\hline
Type           & $Pe$  & $a$  & $\omega$ & $\Psi$ & $e_{avg}$ & $e_{max}$ \\ \hline
Training       & 1000  & 0.35 & 8        & 234.5 & 6.03E-03  & 3.92E-02  \\
Training       & 2000  & 0.35 & 6        & 1667.4 & 1.08E-02  & 3.27E-02  \\
Training       & 5000  & 0.45 & 8        & 16,018  & 8.67E-03  & 2.48E-02  \\
Training       & 20,000 & 0.45 & 12       & 113,910 & 8.12E-03  & 5.71E-02  \\
Predictive (a) & 1000  & 0.45 & 12       & 284.8 & 4.10E-02  & 6.07E-02  \\
Predictive (b) & 5000  & 0.35 & 12       & 2605.3 & 1.10E-02  & 3.60E-02  \\
Predictive (c) & 5000  & 0.35 & 6        & 10,421  & 1.64E-02  & 2.79E-02  \\
Predictive (d) & 10,000 & 0.45 & 8        & 64,072  & 1.26E-02  & 5.47E-02  \\
Predictive (e) & 20,000 & 0.35 & 8        & 93,789  & 3.03E-02  & 5.41E-02  \\ \hline
\end{tabular}
\end{table}

\begin{figure} [H]
     \centering
     \begin{subfigure}[b]{0.49\textwidth}
         \centering
         \includegraphics[width=\textwidth]{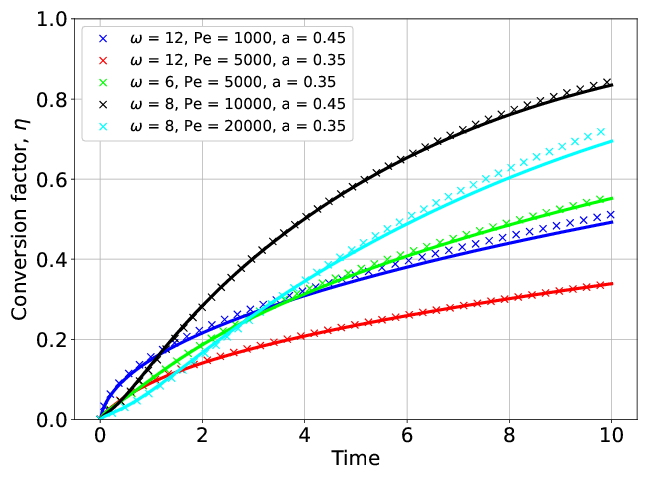}
         \caption{}
     \end{subfigure}
     \hfill
     \begin{subfigure}[b]{0.49\textwidth}
         \centering
         \includegraphics[width=\textwidth]{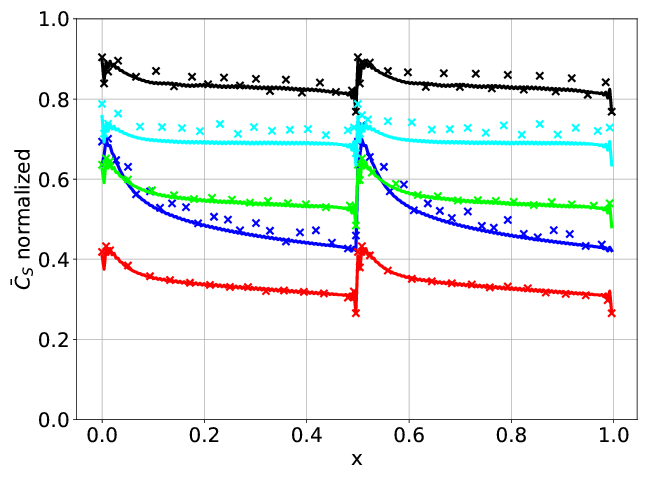}
         \caption{}
     \end{subfigure}
     \hfill
     \begin{subfigure}[b]{0.49\textwidth}
         \centering
         \includegraphics[width=\textwidth]{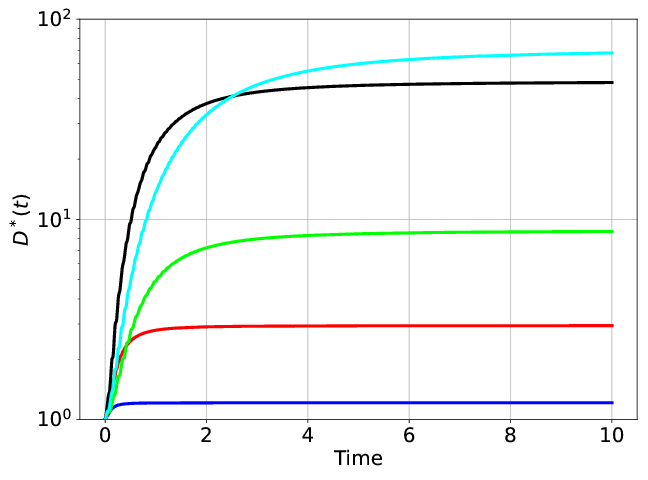}
         \caption{}
     \end{subfigure}
      \caption{(a) Conversion factor for the predictive cases from Table~\ref{tableABC}; (b) normalized $\bar{C}_s$ at $t = 10$; (c) $D^*(t)$ in the upscaled representation for the ABC velocity field. The $\times$’s depict results from the reference 2D models and the solid curves show the upscaled 1D representation. Legend in (a) also applies to (b) and (c).}
        \label{fig:convFactor_Dstar_ABC}
\end{figure}

Figure~\ref{fig:convFactor_Dstar_ABC}a and b show the conversion factor and normalized $\bar{C}_s$ for the five predictive runs. Reasonable agreement between the 2D and 1D representations is obtained, though some discrepancies are evident in the reaction locations in Figure~\ref{fig:convFactor_Dstar_ABC}b. More specifically, the 1D representation is not able to capture the high-frequency oscillations in the spatial distribution of the chemical reaction ($\bar{C}_s$), though it does capture the general trends, with slight shifts in two of the cases. Figure~\ref{fig:convFactor_Dstar_ABC}c shows that $D^*(t)$ in the upscaled representation again extends over a large range, with values approaching 100.

\subsubsection*{ABC velocity field with $N_f = 4$}

\begin{figure} [H]
    \centering
    \includegraphics[width=0.99\textwidth,trim = 70 0 30 0,clip]{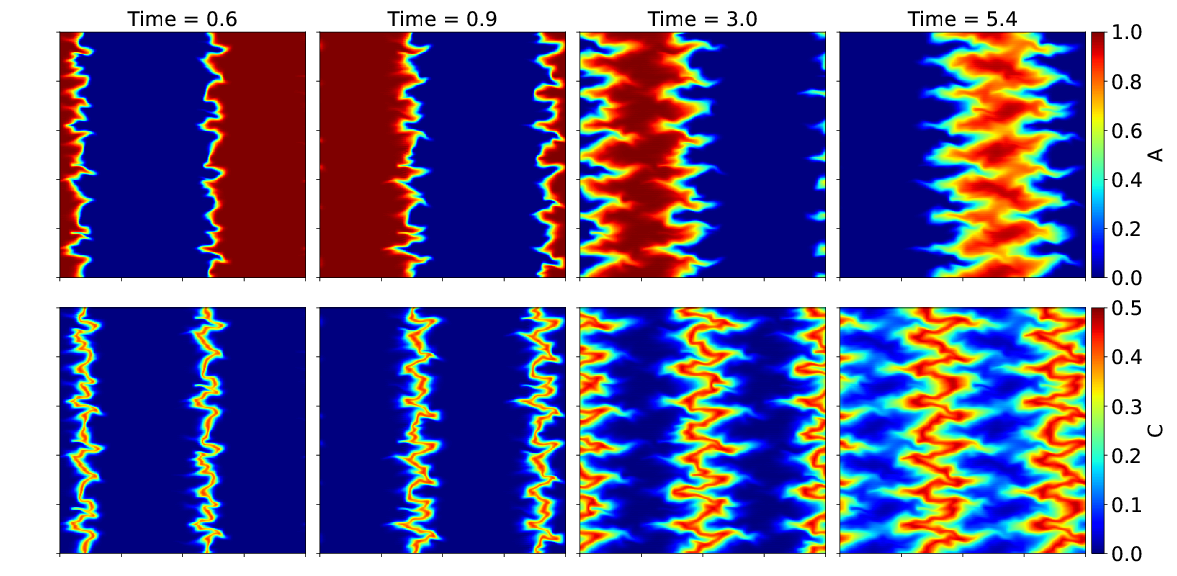}
    \caption{Concentration of component $A$ (top row) and component $C$ (bottom row) at $t$ = 0.6, 0.9, 3.0 and 5.4 for $Pe = 5000$, $\boldsymbol{\omega} = [4,7,11,15]$ and $a = 0.3$ for ABC velocity field with $N_f=4$.}
    \label{fig:maps2D_Pe5000_ws4_7_11_15_a030_ABC_v2}
\end{figure}

To explore the behavior of the upscaled model with multiple frequencies, we now consider the ABC field with $N_f = 4$ and $\boldsymbol{\omega}=[\omega_1,\omega_2,\omega_3,\omega_4]=[4,7,11,15]$, $\boldsymbol{\gamma}=[\gamma_1,\gamma_2,\gamma_3,\gamma_4]=[0,0.3,0.6,0.9]$, with each frequency having amplitude $a=0.30$. Concentration results from reference 2D solutions for this case are shown in Figure~\ref{fig:maps2D_Pe5000_ws4_7_11_15_a030_ABC_v2}. The presence of small-scale features associated with the higher frequencies are evident, especially at $t = 0.6$ and $t = 0.9$. At later times, many of the high-frequency features have coalesced due to diffusion, and only the lower-frequency features are evident. 

\begin{figure} [H]
    \centering
    \includegraphics[width=0.49\textwidth,trim = 0 0 0 0,clip]{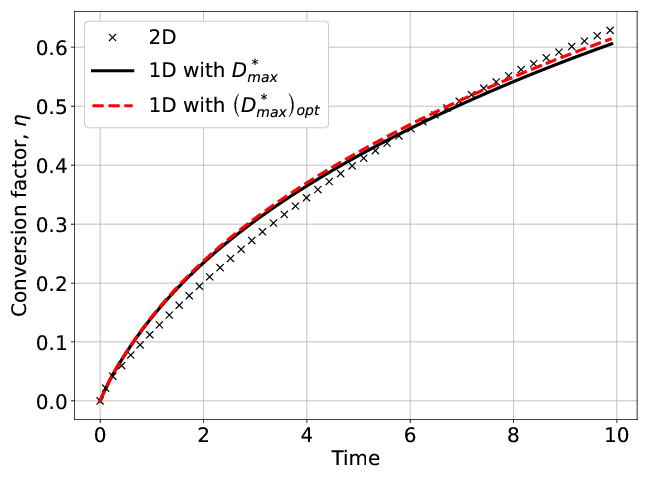}
    \caption{Conversion factor for ABC velocity field with $N_f = 4$. Black $\times$’s indicate the reference 2D result; black solid curve is constructed using $\Dsm$ from Eq.~\ref{eq:DstarMax_fit} with $c=0.000746$ and $\omega = 4$; red dashed curve is constructed using $\Dsmopt$ from minimization problem (Eq.~\ref{eq:invProblem}). Both 1D results use $l(t)$ from particle tracking for ABC velocity field with $N_f = 4$.}
    \label{fig:convFactorABC_ws4_7_11_15}
\end{figure}

The upscaled model relies on tracking $l(t)$ under pure advection. Thus both small and large-scale features affect $l(t)$ at all times. The $D^*(t)$ in Eq.~\ref{eq:Dstar} was not formulated to approximate asymptotic limits with two or more frequencies, so we would not expect the upscaled model to fully capture the complex behaviors associated with multiple frequencies over the full time frame. Reasonable predictions can be obtained, however, by recognizing that the largest-scale features control the diffusion time, and then estimating $\Dsm$ in Eq.~\eqref{eq:DstarMax_fit} based on the smallest frequency involved ($\omega=4$). Results using this approach are shown in Figure~\ref{fig:convFactorABC_ws4_7_11_15} as the solid black curve. It is evident that this curve closely approximates the 2D results ($\times$'s), though it does slightly overestimate the conversion factor at early times and underestimate at long times. The 1D model constructed using $\Dsmopt$ (via Eq.~\ref{eq:invProblem}), shown as the red dashed curve, provides a very similar result. We note finally that it may be possible to generalize the form of $D^*(t)$ (Eq.~\ref{eq:Dstar}) to account for multiple frequencies, using, e.g., a superposition-like representation. This would, however, entail the specification of more than one $\Dsm$ parameter.

\section{Concluding remarks} \label{sec:conclusions}

In this work we introduced a new upscaled model able to represent, in a 1D Eulerian setting, mixing-limited bimolecular reactions driven by variable 2D velocity fields. The key upscaled quantity in the 1D model is the time-dependent effective dispersion. To account for flow kinematics, i.e., the enhanced mixing caused by the stretching of the front due to the variable velocity field, we connect the early-time behavior of the dispersion to the length that the mixing interface would experience under purely advective conditions. This length can be readily computed for general velocity fields with particle tracking, thus providing a computationally efficient way of incorporating essential information from the 2D flow field into the 1D representation. At long times diffusion will counteract interface stretching, and our model accounts for this by introducing an asymptotic limit for the dispersion. This quantity, which represents the only free parameter in the model, is estimated using scaling arguments after calibration for a few reference cases. The governing reactive transport equations (in both 1D and 2D) are solved using a pseudo-spectral method that provides converged results at reasonable computational cost.

Detailed results were presented for 2D oscillatory and ABC velocity fields. We demonstrated that the upscaled model provides very accurate representations of the reaction conversion factor, and reasonable approximations of the spatial distribution of the chemical reaction, over a wide range of Peclet numbers and velocity-field parameters. The high level of agreement observed provides validation of the scaling arguments used in constructing the upscaled representation. It is evident from the numerical results that the early-time approximation (Eq.~\ref{eq:Dstar_noDiff}) indeed provides accurate predictions for cases where the mixing front experiences very substantial stretching (see, e.g., Figure~\ref{fig:maps2D_Pe40000_w4_a0.5_Oscil} and Figure~\ref{fig:convFactor_Dstar_oscil}). We also showed that our model is applicable for non-reacting systems, and that results from this case can be used in upscaled reactive transport models. The accuracy of the upscaled model for the four-frequency ABC velocity example suggests that the model can be generalized to treat a wide range of flow fields.

There are many promising directions for future research in this area. In this work, our goal was the representation of mixing (and reaction) rates rather than spreading. It will be of interest to develop consistent two-equation representations able to treat both effects in 1D models. Restrictions resulting from our use of periodic boundary conditions did not allow us to explore model behavior for arbitrarily large plume stretching. Further investigation of the limits of applicability of the model, particularly the early-time approximation, should thus be pursued. It may also be useful to generalize the form of the time-dependent dispersion expression (i.e., consider alternatives to Eq.~\eqref{eq:Dstar}), especially when multiple frequencies are involved. This could lead to treatments that are applicable for general velocity fields. It is also possible that further computational savings could be achieved using analytical or hybrid solutions of the 1D equations, and this should be considered. Finally, we note that many of the ideas and treatments presented here may be applicable to upscaling from the pore to Darcy scale. Research along these lines should be pursued. 

\section*{Acknowledgments}
The first author is grateful to Petrobras for financial support. We also wish to thank Hamdi Tchelepi for useful discussions. 
\pagebreak
\begin{appendices}
\appendix
\counterwithin{figure}{section}
\section{Numerical convergence analysis}

The high-resolution solutions we construct, both for Eq.~\ref{eq:conserv_adim} (2D system) and Eq.~\ref{eq:conserv_adim_upsc} (1D system), would be very costly to obtain with conventional finite-volume or finite-difference methods. For this reason, as noted in the main text, we apply a pseudo-spectral method that is able to resolve the sharp fronts associated with high-Peclet-number reactive transport solutions. The method, described in detail by \cite{adrover2002spectral}, performs numerical integration with a fourth-order Runge-Kutta scheme and applies a Fourier approximation of the spatial derivatives. Spectral convergence is observed and efficiency is achieved through the use of FFT computations. The time step size can be limited by either advective or diffusive terms, and it is determined in this work through a consideration of both effects. In the solutions presented in Section~\ref{sec:applications}, we use between 4700 and 33,000 time steps to integrate from the initial state to $t_f = 10$.

\begin{figure} [H]
    \centering
    \includegraphics[width=0.49\textwidth,trim = 0 0 0 0,clip]{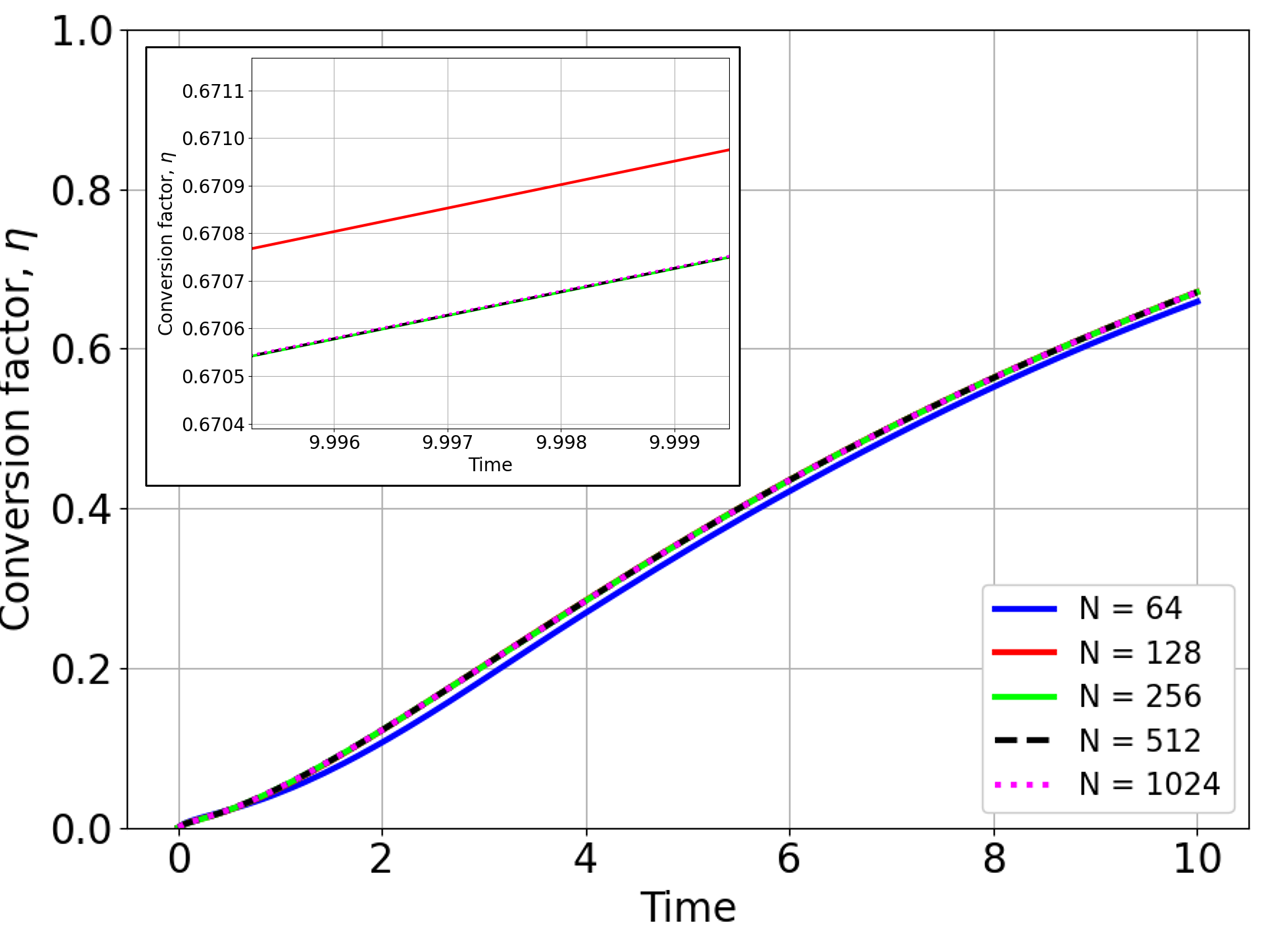}
    \caption{Convergence results using the pseudo-spectral method for 2D solutions with varying numbers of grid points ($N=N_x=N_y$) for the oscillatory velocity field with $Pe = 40,000$, $\omega = 20$, and $a = 0.05$. The inset shows the solution near $t=10$.}
    \label{fig:convergOscil}
\end{figure}

The most difficult solutions correspond to Eq.~\ref{eq:conserv_adim} with large values of $Pe$, $\omega$, and $a$, as these specifications result in sharp concentration gradients that persist in time. Convergence results for conversion factor, for the oscillatory velocity field with $Pe = 40,000$, $\omega = 20$, and $a = 0.05$, are presented in Figure~\ref{fig:convergOscil}. Two-dimensional solutions are shown with different numbers of grid points ($N$) in each coordinate direction, with $N=N_x=N_y$ ranging from 64 to 1024. The inset shows results toward the end of the run. A small offset is visible for $N=64$, though results for $N \geq 128$ essentially collapse. By $N=256$, which is the value used in this study, the results for conversion factor have converged to five significant figures. For the ABC velocity field, convergence is also obtained with $N=256$. In these 2D solutions, this corresponds to a total of $N^2=65,536$ grid points. We note that, for all cases in this study with $N_f = 1$, fewer grid points could be used in the $y$-direction by exploiting solution periodicity (at a scale smaller than $L_y$) and then treating nonsquare domains.

Solutions for the spatial distribution of the reaction throughout the domain (${\bar C}_s$) display oscillations, particularly near $x = 0$ and $x = 0.5$, that reduce with increasing $N$. This occurs because the initial condition is discontinuous, leading to the appearance of the Gibbs phenomenon, described by \cite{canuto2007spectral} in the context of spectral computations. For $N=256$, away from $x = 0$ and $x = 0.5$, these oscillations are very small -- of magnitude 0.003, which is negligible compared to the $O(1)$ ${\bar C}_s$ solution. These oscillations could be reduced by proceeding to higher $N$, though this was not necessary for the cases considered in this study.

\end{appendices}

\pagebreak

\bibliographystyle{apalike}

\bibliography{library}
\end{document}